\documentclass[aps,pra,reprint,showkeys]{revtex4-2}
\usepackage{graphicx}
\usepackage{soul,color}
\sethlcolor{cyan}

\begin{document}

\title{On the possibility of incorporation of the iron impurity into $A$ sites in SrTiO$_3$}

\author{I. A. Sluchinskaya}
\author{A.I. Lebedev}
\email[]{swan@scon155.phys.msu.ru}
\affiliation{Physics Department, M.V. Lomonosov Moscow State University, Moscow, 119991 Russia}

\date{\today}

\begin{abstract}
The local environment and oxidation state of the Fe impurity in strontium titanate
are studied using XAFS spectroscopy. The influence of annealing temperature and
deviation from stoichiometry on the possibility of incorporation of the impurity
into the $A$ and $B$~sites of the perovskite structure is studied. The results
obtained from the X-ray diffraction, XANES spectra, and EXAFS spectra suggest
that at high annealing temperature the iron atoms, at least partially (up to 30\%),
enter the $A$~sites in SrTiO$_3$. The obtained results agree with results of
first-principles calculations, according to which the iron at the $A$~site exhibits
strong off-centering (the displacement of $\sim$1~{\AA}), similar to that previously
established in SrTiO$_3$ samples doped with Mn and Co.

\texttt{DOI: 10.21883/PSS.2022.03.53189.242 (Physics of the Solid State 64, 345 (2022))}
\end{abstract}

\keywords{strontium titanate, iron, doping, XAFS, off-center atoms.}

\maketitle

\section{Introduction}

The search for new multiferroics---multifunctional materials that open up new
possibilities for modern electronics and spintronics---is currently an actual
problem. In particular, a part of this problem concerns the search for new
magnetic off-center impurities, doping with which can induce magnetoelectric
interaction of their dipole and magnetic moments.

Oxides with the perovskite structure with a general formula \emph{AB}O$_3$ have
long been a subject of intensive research due to their unique properties---from
ferroelectricity, piezoelectricity to colossal magnetoresistance and
multiferroicity~\cite{TejucaFierro}. Among these oxides, strontium titanate SrTiO$_3$
is an important material whose electrical, optical, thermoelectric, and superconducting
properties can be (or are already) used in various electronic and other
devices~\cite{ApplPhysLett.73.190,ApplPhysLett.76.1324,JMaterSciMaterElectron.14.483,
ApplPhysLett.87.092108,PhysRevLett.112.207002,ApplPhysLett.105.183103}. The
photocatalytic and sensory properties of doped SrTiO$_3$ are also of
interest~\cite{JPhysChemB.106.5029,JPhysChemB.108.8992,SensActuatorsB.136.489}.

To obtain multiferroic properties, the crystal must simultaneously contain magnetic
and dipole moments. Doping of SrTiO$_3$ with magnetic impurities of transition $3d$
elements enables to create magnetic moments and thereby to expand the functionality
of this material by controlling it with a magnetic field. However, the position
of the impurity atoms in a crystal and their local environment can significantly
influence their magnetic properties. In addition, $3d$ elements are characterized
by the possibility of their existence in several oxidation states, which depend on
the position and local environment of these atoms as well as on the presence of
other donors and acceptors in the samples. Therefore, when studying such doped
samples, it is necessary to study the local environment of the $3d$ elements.

To create dipole moments, one can try to use the idea of an off-center impurity.
Off-centering is possible if a substitutional impurity enters the $A$~site of the
perovskite structure since the ionic radii of all $3d$ elements are much smaller
than that of strontium. In this work, we analyze the possibility of Fe to enter
the $A$~site and to displace into an off-center position.

Strontium titanate doped with $3d$ elements has been studied since the 1960s. For
a long time, it was believed that $3d$ elements in SrTiO$_3$ replace only
Ti atoms due to the closeness of their ionic radii. The first results indicating
the possibility of incorporation of $3d$ elements into the $A$~site were
obtained for the manganese impurity. These studies began with the discovery of
unusual relaxations of the dielectric constant in SrTiO$_3$(Mn)
samples~\cite{PhysSolidState.46.1442,JApplPhys.98.056102}.
The study of the conditions for the appearance of these relaxations first led
to the assumption~\cite{ApplPhysLett.86.172902} and then to a direct proof,
using the EXAFS spectroscopy, of the incorporation of manganese into the
$A$~site~\cite{JETPLett.89.457,ApplPhysLett.96.052904,BullRASPhys.74.1235,
AdvFunctMater.22.2114,ChemMater.32.4651}. The technological conditions for this
were a deviation from stoichiometry toward titanium and a high annealing
temperature~\cite{ApplPhysLett.86.172902,JETPLett.89.457,ApplPhysLett.96.052904,
BullRASPhys.74.1235,AdvFunctMater.22.2114}. Heat treatment in a reducing atmosphere
also promoted the incorporation of manganese into the
$A$~sites~\cite{AdvFunctMater.22.2114,ChemMater.32.4651}. The interest to Mn-doped
strontium titanate was stimulated by publications that revealed the appearance of
multiferroic properties in it~\cite{PhysRevLett.101.165704,JPhysCondensMatter.20.434216}.

Somewhat later, using a similar approach to the synthesis of samples and the
EXAFS spectroscopy technique, we prepared and studied SrTiO$_3$ samples with an
off-center cobalt impurity~\cite{PhysSolidState.61.390,JAlloysComp.820.153243}.

Our interest to the samples doped with Fe is caused by the fact that dielectric
measurements on ceramic SrTiO$_3$(Fe) samples with an intentionally created
strontium deficiency~\cite{PhysRevMater.2.094409} revealed similar relaxation
phenomena as in samples doped with off-center Mn and Co impurities. This could
indicate the appearance of off-center iron atoms.

A large number of papers have been devoted to studies of iron-doped strontium
titanate. It was believed that these samples are solid solutions between
strontium ferrite SrFeO$_3$ and strontium titanate SrTiO$_3$, that is, the
substitution occurs at the $B$~site. Due to the difference in the oxidation
states of titanium ions (Ti$^{4+}$) and a part of Fe ions that are in the
Fe$^{3+}$ state, the oxygen vacancies are usually present in these samples to
maintain their electrical neutrality. This doping feature and the presence
of iron atoms in two oxidation states (Fe$^{4+}$ and Fe$^{3+}$) in samples
synthesized in air results in a mixed (electronic and ionic) conductivity, which
is important for the development of solid electrolytes, fuel cells, electrochemical
oxygen sensors, and catalysts~\cite{JElectroceram.28.80}. The study of
the photochromic effect in iron-doped strontium titanate~\cite{PhysRevB.4.3623}
and consequent finding ways to control this effect using an electric
field~\cite{PhysChemChemPhys.13.20779} made this material a candidate for its
use in resistive random access memory (ReRAM) devices.

The SrTiO$_3$(Fe) samples were studied in detail by XAFS spectroscopy. In
Refs.~\cite{PhysChemChemPhys.13.20779,PhysRevB.76.174107}, using XAFS spectroscopy,
EPR, Raman scattering, and theoretical calculations, the ratio of the number
of iron atoms in the +4 and +3 oxidation states as well as the ratio of the
number of isolated Fe$^{3+}$ ions and their complexes with the nearest oxygen
vacancies were estimated in samples prepared under various conditions. In these
papers, the interatomic distances in the local environment of Fe atoms were also
determined. A slight increase in the Debye--Waller factor for the Fe$^{4+}$
ion was explained by the Jahn--Teller effect, which was later confirmed by
\emph{ab initio} calculations~\cite{PhysRevB.77.075111}. An analysis of the
X-ray absorption near-edge structure (XANES) in spectra of thin SrTiO$_3$(Fe)
films prepared by pulsed laser deposition~\cite{PhysChemChemPhys.15.8311} in
the vicinity of the Fe $L$-edge made it possible to establish that iron in these
samples is in a mixture of +2 and +3 oxidation states. Annealing the films in vacuum
resulted in an increase in the Fe$^{2+}$/Fe$^{3+}$ ratio. The contrast between
the results obtained for bulk samples and thin films was explained by a specific
defect structure of thin films caused by their preparation under nonequilibrium
conditions.

To check the possibility of the incorporation of Fe impurities into the $A$~sites
of strontium titanate, we used the same synthesis conditions as were used before
in our studies of off-center Mn and Co impurities in SrTiO$_3$. The obtained
samples were studied by XAFS spectroscopy. The experimental results were analyzed
taking into account the results of first-principles calculations, in which
theoretical models of impurity centers we could meet in the experiment according
to the literature data were constructed.

\section{Experimental and calculation techniques}

SrTiO$_3$ samples doped with Fe with an impurity concentration of 2--3\%
and different deviation from stoichiometry were prepared by solid-phase
synthesis. The starting components were SrCO$_3$, nanocrystalline TiO$_2$
obtained by hydrolysis of tetrapropylorthotitanate and dried at 500$^\circ$C,
and Fe$_2$O$_3$. The components were weighed in the required proportions,
ground in acetone and annealed in alumina crucibles in air at 1100$^\circ$C
for 8~hours. The resulting powders were ground and annealed again under
the same conditions. Several samples were additionally annealed in air at
1500$^\circ$C for 2~hours. To incorporate impurities into the $A$ or $B$~sites
of the perovskite structure, the composition of the samples was deliberately
deviated from stoichiometry, respectively, toward excess titanium or strontium.

X-ray absorption spectra in the regions of extended fine structure (EXAFS)
and near-edge structure (XANES) were recorded by detecting X-ray fluorescence
at the Fe $K$-edge (7.112~keV) at 300 K at the KMC-2 station of the BESSY
synchrotron radiation source (electron energy 1.7~GeV, maximum beam current
290~mA). The incident radiation was monochromatized with a (111)-oriented
Si$_{1-x}$Ge$_x$ double-crystal monochromator. The intensity of this radiation
($I_0$) was measured using an ionization chamber. The fluorescence intensity $I_f$
was measured with a R{\"O}NTEC silicon drift detector operating in energy dispersive
mode. The powders were rubbed into the surface of adhesive tape, which was then
folded to obtain the optimum sample thickness.

The extraction of the oscillating EXAFS function $\chi(k)$ from the fluorescence
excitation spectra $\mu(E) = I_f / I_0$ (here $E$ is the X-ray photon energy)
was carried out in the traditional way~\cite{BullRASPhys.60.1533,PhysRevB.55.14770}.
After subtracting the background below the absorption edge, the monotonic part
$\mu_0(E)$ of the spectrum was approximated with splines, and the dependence
$\chi = (\mu - \mu_0) / \mu_0$ was calculated as a function of the photoelectron
wave vector $k = (2m(E - E_0)/\hbar^2)^{1/2}$. The energy corresponding to the
inflection point on the $\mu(E)$ curve was taken as the photoelectron energy
origin $E_0$. The EXAFS spectra were processed with a widely used IFEFFIT software
package~\cite{JSynchrotronRad.12.537} similarly to~\cite{PhysSolidState.61.390,PhysSolidState.59.1512}.
For each sample, 3--4 spectra were recorded, they were then independently processed,
and the obtained $\chi(k)$ curves were averaged. Details of data processing are
given in Ref.~\cite{PhysRevB.55.14770}.

The geometry, electronic structure, and magnetic moment of impurity centers were
calculated from first principles using the ABINIT software package in the LDA+$U$
approximation. To correctly describe Fe atoms with a partially filled $d$ shell,
we used the PAW pseudopotentials~\cite{ComputMaterSci.81.446}. The parameters
describing the Coulomb and exchange interactions for the impurity atom were
$U = 5$~eV and $J = 0.9$~eV. As shown in our earlier studies~\cite{Ferroelectrics.501.1},
in order to obtain relatively narrow impurity bands in the energy spectrum, it is
necessary to use supercells containing more than 40~atoms. Therefore, to simulate
isolated impurity centers, we used 80-atom FCC supercells in which one of Ti$^{4+}$
ions at the $B$~site or Sr$^{2+}$ ions at the $A$~site was replaced by an impurity
atom. The cutoff energy of plane waves in the calculations was 30~Ha (816~eV), and
the integration over the Brillouin zone was carried out on the
$2\sqrt{3} \times 2\sqrt{3} \times 2\sqrt{3}$ Monkhorst--Pack mesh. The relaxation
of the lattice parameters and atomic positions was carried out until the forces
acting on the atoms became less than 10$^{-5}$~Ha/Bohr (0.5~meV/{\AA}). When
a change in the oxidation state of iron was necessary, oxygen vacancies as well as
donor or acceptor impurities located at the maximum distance from the
impurity atom were added to the structure.

\begin{table*}
\caption{\label{table1}Local environment of Fe ions in different theoretical models.}
\begin{ruledtabular}
\begin{tabular}{ccccccccc}
 &  & \multicolumn{7}{c}{Shell} \\
\cline{3-9}
\smash{\raisebox{7pt}{Model}}    & \smash{\raisebox{7pt}{$S$}} & 1 & 2 & 3 & 4 & 5 & 6 & 7 \\
\hline
Off-center Fe$^{2+}$ ion at the  & 2   & 2.027 & 2.963 & 3.048 & 3.170 & 3.633 & 3.818 & 4.021 \\
$A$~site, displacement [100]     &     & (4O)  & (4Ti) & (1Sr) & (4O)  & (4O)  & (4O)  & (4Ti) \\

Off-center Fe$^{2+}$ ion at the  & 2   & 2.104 & 2.156 & 2.815 & 3.195 & 3.377 & 3.434 & 3.437 \\
$A$~site, displacement [110]     &     & (4O)  & (1O)  & (2Ti) & (2O)  & (2Sr) & (4O)  & (4Ti) \\

Off-center Fe$^{2+}$ ion at the  & 2   & 2.013 & 2.607 & 2.936 & 3.233 & 3.514 & 3.642 & 3.708 \\
$A$~site, displacement [111]     &     & (3O)  & (1Ti) & (6O)  & (3Ti) & (3Sr) & (3O)  & (3Ti) \\

Isolated Fe$^{3+}$ ion at the $B$~site & 5/2   & 1.988 & 3.374 & 3.904 & 4.368 & 5.521  & \\
(compensated by Y)                     &       & (6O)  & (7Sr) & (6Ti) & (24O) & (12Ti) & \\

Fe$^{3+}$--$V_{\rm O}$ complex   & 5/2 & 1.902 & 1.964 & 3.416 & 3.499 & 3.702 & 3.926 & 4.149 \\
(compensated by Sc)              &     & (1O)  & (4O)  & (4Sr) & (4Sr) & (1Ti) & (4Ti) & (1Ti) \\

Isolated Fe$^{4+}$ ion at the $B$~site & 2 & 1.932 & 3.360 & 3.890 & 4.358 & 5.515  & \\
                                       &   & (6O)  & (8Sr) & (6Ti) & (24O) & (12Ti) & \\
\end{tabular}
\end{ruledtabular}
\end{table*}

\section{Results of first-principles calculations}
\label{Sec3}

For reliable identification of defects in SrTiO$_3$(Fe) samples by EXAFS
spectroscopy, we need structural models of possible defects. For this purpose,
calculations of the local environment, electronic structure, and magnetic moment
of several iron-containing complexes were carried out. According to the literature,
in bulk SrTiO$_3$ samples, in addition to iron atoms at the $B$~sites in the
+4 oxidation state, a certain fraction of Fe atoms at these sites can be in the
+3 oxidation state both as isolated ions and complexes with oxygen vacancies.
Therefore, we need to study the structure of possible complexes of Fe with oxygen
vacancies. For atoms at the $A$~site, the supposed oxidation state is +2. Such
doping does not violate the electrical neutrality of the samples, and it is sufficient
to consider only point defects (isolated impurity atoms) in them. The results of
calculations of the local environment and magnetic moments of the ions are
given in Table~\ref{table1}.

Calculations for an isolated Fe$^{3+}$ ion at the $B$~site, for which a donor Y atom
at the Sr site was added to the supercell to obtain the required oxidation state,
showed that the energetically most favorable state of the ion is the high-spin
state with a magnetic moment of $5\mu_B$. In the case of the Fe$^{3+}$--$V_{\rm O}$
complex with the nearest oxygen vacancy and a compensating acceptor impurity
Sc$_{\rm Ti}$, the magnetic moment of the ion is also $5\mu_B$. For an isolated
Fe$^{4+}$ ion, the magnetic moment is $4\mu_B$.

For the Fe$^{2+}$ ion at the $A$~site, the on-center position of the impurity atom
turned out to be unstable. That is why we calculated three configurations corresponding
to possible off-center displacements of this atom along the [100], [110], and
[111] directions. The values of the corresponding off-center displacements were
1.14, 0.96, and 0.95~{\AA}. In all three configurations, the Fe ion is in a
high-spin state (magnetic moment $4\mu_B$).

In the obtained results, attention should be drawn to the fact that the Fe--O
interatomic distances for off-center atoms at the $A$~site are close to the Fe--O
distances for the $B$~site (Table~\ref{table1}). This can strongly complicate the
analysis of the EXAFS spectra.

\section{Results and discussion}

\subsection{X-ray diffraction}

In what follows, to designate the samples we will use names consisting of a
number---the nominal concentration of the impurity in percent, a letter---a
site into which the impurity should enter, and numbers---the annealing
temperature (for example, 2A1500 means the Sr$_{0.98}$Fe$_{0.02}$TiO$_3$ sample
annealed at 1500$^\circ$C).

\begin{figure}
\includegraphics{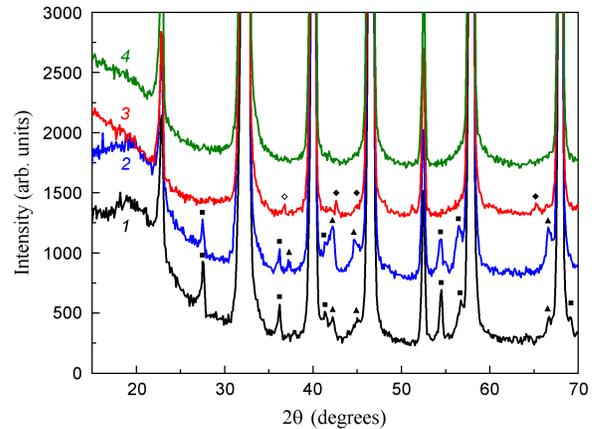}
\caption{\label{fig1}X-ray diffraction patterns of SrTiO$_3$(Fe) samples: (1) 2A1100,
(2) 3B1100, (3) 2A1500, and (4) 3B1500. The squares denote the reflexes of the
TiO$_2$ second phase, and the triangles denote the position of the lines of the
Sr$_3$Ti$_2$O$_7$ Ruddlesden--Popper phase. The filled diamonds mark the
Sr$_3$Fe$_2$O$_7$ lines, and an open diamond indicates the line of the Fe$_2$TiO$_5$
phase.}
\end{figure}

X-ray studies of SrTiO$_3$(Fe) samples showed that all samples have a cubic
perovskite structure at 300~K (Fig.~\ref{fig1}). Among the studied samples, only
one sample turned out to be single-phase---the 3B1500 sample. The diffraction
patterns of the samples annealed at 1100$^\circ$C clearly showed the lines of TiO$_2$
and of the Ruddlesden--Popper Sr$_3$Ti$_2$O$_7$ phase, whose concentrations,
however, were not high. The appearance of the Ruddlesden--Popper phase, which contains
an excess of strontium, compensates for the segregation of titanium in the form
of TiO$_2$. The ratio of intensities of the corresponding lines is consistent
with the deviation from stoichiometry created in the samples. The diffraction
pattern of the 2A1500 sample showed weak lines,
three of which can be identified as the Ruddlesden--Popper phase Sr$_3$Fe$_2$O$_7$,
and the line at $2\theta = 36.75^\circ$---as the precipitation of Fe$_2$TiO$_5$.
The line at $2\theta = 51.3^\circ$ remains unidentified. A very small amount
of the second phase in the 2A1500 sample may indicate successful incorporation
of the iron atoms into the $A$~sites.

\subsection{XANES Spectra}

\begin{figure}
\includegraphics{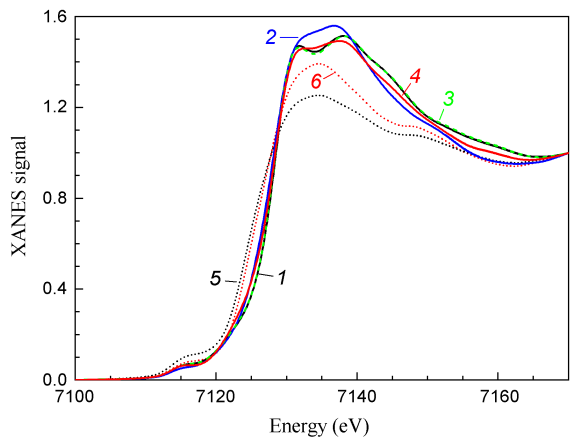}
\caption{\label{fig2}XANES spectra of SrTiO$_3$(Fe) samples and reference
compounds: (1) 2A1100, (2) 2A1500, (3) 3B1100, (4) 3B1500, (5) FeTiO$_3$,
(6) Fe$_2$O$_3$.}
\end{figure}

To determine the oxidation state of the Fe impurities in SrTiO$_3$, the position of
the absorption edges in the XANES spectra of the SrTiO$_3$(Fe) samples was compared
with the position of the absorption edges in the FeTiO$_3$ and Fe$_2$O$_3$ reference
compounds (Fig.~\ref{fig2}). Unfortunately, the above method for determining
the oxidation state was not very successful. The shift of the absorption edge between
the reference samples of divalent and trivalent Fe as well as between the samples
obtained under different preparation conditions, was only $\sim$0.5~eV. This was
insufficient for reliable identification of the impurity oxidation state. The weak
influence of the ionic charge on the position of the Fe $K$-edge was also noticed
in Ref.~\cite{PhysChemChemPhys.13.20779}.

It is noteworthy that the shape of the XANES spectra of both samples annealed at
1500$^\circ$C in the energy range 7131--7142~eV differs from that of the samples
annealed at 1100$^\circ$C. In our opinion, this may indicate a change in the
position of the impurity and the incorporation of some iron atoms into the $A$~sites.

\subsection{EXAFS Spectra}

To determine the structural position of the Fe impurities in SrTiO$_3$, we analyzed
the EXAFS spectra. When analyzing the data, we took into account the lattice parameter
obtained from the X-ray measurements, and, when choosing the structural models, the
results of calculations of the defects' geometries which included all relaxations of
the surrounding atoms (Sec.~\ref{Sec3}). In addition, the possibility of the
impurity atoms to enter both the $A$ and $B$~sites of the perovskite structure was
considered~\cite{PhysSolidState.61.390,JAlloysComp.820.153243}.

\begin{figure}
\includegraphics{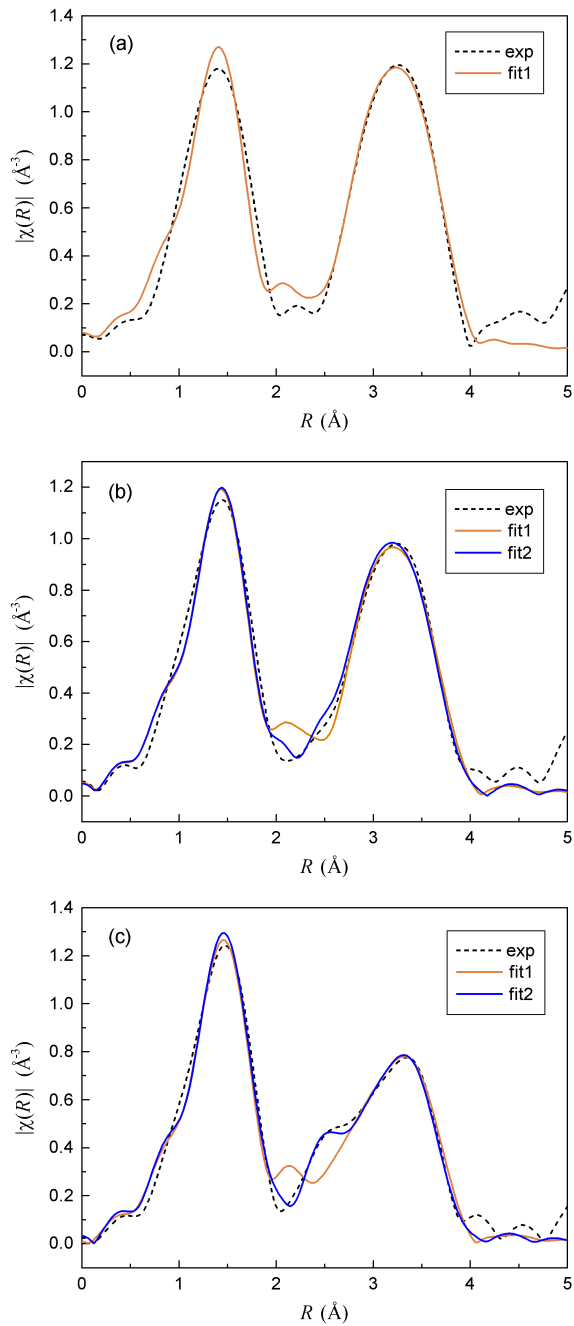}
\caption{\label{fig3}Fourier transforms of experimental EXAFS spectra in the $R$-representation
(exp) and their best fits in the model with Fe atoms only at the $B$~sites (fit1) and
the model with the simultaneous presence of iron at the $A$ and $B$~sites (fit2).}
\end{figure}

The results of processing of the experimental EXAFS data of SrTiO$_3$(Fe) samples and
their best fits using different theoretical models are shown in Fig.~\ref{fig3}.
The used models were the model with the Fe atom at the $B$~site and the model with the
simultaneous incorporation of the impurity into the $B$ and $A$~sites. In the latter
case, the atom at the $A$~site was assumed to be off-center.

An analysis of the EXAFS data showed the following. In the EXAFS region, regardless
of the preparation conditions, the spectra were qualitatively described by a model
in which the Fe atom replaces the atom at the $B$~site with the interatomic Fe--O distances
varying from 1.906~{\AA} for the 3B1100 sample to 1.930~{\AA} for the 3B1500 sample,
and to 1.949~{\AA} for the 2A1500 sample. Taking into account the results obtained
in Ref.~\cite{PhysChemChemPhys.13.20779}, we suggest that the iron atoms are
predominantly in the +4 oxidation state in samples annealed at 1100$^\circ$C,
and in the +3 and +4 oxidation states in samples annealed at 1500$^\circ$C. The
obtained interatomic distances agree with the results of first-principles calculations
(Table~\ref{table1}). However, the difference in interatomic distances in the samples,
which were annealed at the same temperature (1500$^\circ$C) but had different
deviation from stoichiometry, when taking into account the results of calculations
in Sec.~\ref{Sec3}, enable us to suppose that a part of the iron atoms enter the
$A$~site since the average Fe--O distance for the $A$~site is bigger than for
the $B$~site.

As follows from Fig.~\ref{fig3}, the EXAFS spectra of the samples annealed at
1500$^\circ$C having different deviation from stoichiometry are qualitatively different.
The deviation of the curves is especially noticeable in the range 1.9--2.9~{\AA}, in
which additional features associated with a change in the local environment should
appear when the impurity atoms enter the $A$~site (Table~\ref{table1}). In order to
test this hypothesis, we analyzed the EXAFS spectra using a model of simultaneous
incorporation of Fe into both lattice sites, assuming that the direction of the
off-center displacement of the atom is [100]. For this model, the agreement of the
spectra has indeed noticeably improved (Fig.~\ref{fig3}).

The strongest contributions to the EXAFS spectra from the Fe atom at the $A$~site
come from the nearest O and Ti atoms. From the comparison of the Fe--Ti distance of
2.99~{\AA} obtained from the EXAFS data analysis with the distances calculated for
different off-centering models (Table~\ref{table1}) it follows that the most probable
direction of the displacement of the Fe atom from the $A$~site is indeed the [100] one.

According to our data, the fraction of the Fe atoms at the $A$~sites increased from
$\sim$0.24 in the 3B1500 sample to $\sim$0.31 in the 2A1500 sample. It is interesting
that according to our results, at a high annealing temperature some Fe atoms
enter the $A$~sites even for deviation from stoichiometry toward strontium. We
believe that small precipitates of the second phase observed in the X-ray diffraction
of the 2A1500 sample cannot be the cause of such noticeable changes in the EXAFS
spectra.

The overestimated Debye--Waller factors for the nearest oxygen atoms in our samples
(0.008~{\AA}$^2$) are in good agreement with the data for oxidized samples obtained
in Ref.~\cite{PhysRevB.76.174107}. In that work, this overestimation was explained
by the Jahn--Teller distortion around the Fe$^{4+}$ ion. However, our results admit
another explanation---the coexistence of two structural positions of the iron atoms,
which slightly differ in the Fe--O distance.

The low concentration of iron at the $A$~site and the closeness of the Fe--O distances
in models with the atoms at $A$ and $B$~sites did not enable us to more accurately
determine the parameters of structural models. To solve this problem, it is desirable
to synthesize samples with a higher fraction of iron at the $A$~site (for example, by
synthesizing samples at higher annealing temperatures and, possibly, in a reducing
atmosphere) and to measure the EXAFS spectra at lower temperatures.

\section{Conclusion}

In this work, an attempt has been undertaken to incorporate the iron impurity into the
$A$~sites of strontium titanate. The results obtained from the analysis of X-ray
diffraction, XANES spectra, and EXAFS spectra give substantial grounds to suppose that
at a high annealing temperature the Fe atoms enter, at least partially, the $A$~sites.
For more reliable identification of iron at the $A$~site, it is desirable to study
samples with a higher fraction of the atoms at the $A$~site and to analyze the EXAFS
spectra obtained at a lower temperature. According to our data, the Fe impurity in
SrTiO$_3$ exhibits a strong off-centering at the $A$~site (the displacement of
$\sim$1~{\AA}) similar to that previously established for strontium titanate doped
with Mn and Co. This explains the cause of dielectric relaxations in SrTiO$_3$(Fe)
samples observed in Ref.~\cite{PhysRevMater.2.094409}.

\begin{acknowledgments}
The authors express their gratitude to BESSY for the opportunity to carry out our
experiments and financial support of this project.
\end{acknowledgments}


\begin{thebibliography}{35}%
\makeatletter
\providecommand \@ifxundefined [1]{%
 \@ifx{#1\undefined}
}%
\providecommand \@ifnum [1]{%
 \ifnum #1\expandafter \@firstoftwo
 \else \expandafter \@secondoftwo
 \fi
}%
\providecommand \@ifx [1]{%
 \ifx #1\expandafter \@firstoftwo
 \else \expandafter \@secondoftwo
 \fi
}%
\providecommand \natexlab [1]{#1}%
\providecommand \enquote  [1]{``#1''}%
\providecommand \bibnamefont  [1]{#1}%
\providecommand \bibfnamefont [1]{#1}%
\providecommand \citenamefont [1]{#1}%
\providecommand \href@noop [0]{\@secondoftwo}%
\providecommand \href [0]{\begingroup \@sanitize@url \@href}%
\providecommand \@href[1]{\@@startlink{#1}\@@href}%
\providecommand \@@href[1]{\endgroup#1\@@endlink}%
\providecommand \@sanitize@url [0]{\catcode `\\12\catcode `\$12\catcode
  `\&12\catcode `\#12\catcode `\^12\catcode `\_12\catcode `\%12\relax}%
\providecommand \@@startlink[1]{}%
\providecommand \@@endlink[0]{}%
\providecommand \url  [0]{\begingroup\@sanitize@url \@url }%
\providecommand \@url [1]{\endgroup\@href {#1}{\urlprefix }}%
\providecommand \urlprefix  [0]{URL }%
\providecommand \Eprint [0]{\href }%
\providecommand \doibase [0]{https://doi.org/}%
\providecommand \selectlanguage [0]{\@gobble}%
\providecommand \bibinfo  [0]{\@secondoftwo}%
\providecommand \bibfield  [0]{\@secondoftwo}%
\providecommand \translation [1]{[#1]}%
\providecommand \BibitemOpen [0]{}%
\providecommand \bibitemStop [0]{}%
\providecommand \bibitemNoStop [0]{.\EOS\space}%
\providecommand \EOS [0]{\spacefactor3000\relax}%
\providecommand \BibitemShut  [1]{\csname bibitem#1\endcsname}%
\let\auto@bib@innerbib\@empty
\bibitem [{\citenamefont {Tejuca}\ and\ \citenamefont
  {Fierro}(1992)}]{TejucaFierro}%
  \BibitemOpen
  \bibinfo {editor} {\bibfnamefont {L.~G.}\ \bibnamefont {Tejuca}}\ and\
  \bibinfo {editor} {\bibfnamefont {J.~L.~G.}\ \bibnamefont {Fierro}},\ eds.,\
  \href@noop {} {\emph {\bibinfo {title} {Properties and Applications of
  Perovskite-Type Oxides}}}\ (\bibinfo  {publisher} {CRC Press},\ \bibinfo
  {address} {New York},\ \bibinfo {year} {1992})\BibitemShut {NoStop}%
\bibitem [{\citenamefont {Li}\ \emph {et~al.}(1998)\citenamefont {Li},
  \citenamefont {Si}, \citenamefont {West},\ and\ \citenamefont
  {Xi}}]{ApplPhysLett.73.190}%
  \BibitemOpen
  \bibfield  {author} {\bibinfo {author} {\bibfnamefont {H.-C.}\ \bibnamefont
  {Li}}, \bibinfo {author} {\bibfnamefont {W.}~\bibnamefont {Si}}, \bibinfo
  {author} {\bibfnamefont {A.~D.}\ \bibnamefont {West}},\ and\ \bibinfo
  {author} {\bibfnamefont {X.~X.}\ \bibnamefont {Xi}},\ }\bibfield  {title}
  {\bibinfo {title} {Near single crystal-level dielectric loss and nonlinearity
  in pulsed laser deposited SrTiO$_3$ thin films},\ }\href
  {https://doi.org/10.1063/1.121751} {\bibfield  {journal} {\bibinfo  {journal}
  {Appl. Phys. Lett.}\ }\textbf {\bibinfo {volume} {73}},\ \bibinfo {pages}
  {190} (\bibinfo {year} {1998})}\BibitemShut {NoStop}%
\bibitem [{\citenamefont {Eisenbeiser}\ \emph {et~al.}(2000)\citenamefont
  {Eisenbeiser}, \citenamefont {Finder}, \citenamefont {Yu}, \citenamefont
  {Ramdani}, \citenamefont {Curless}, \citenamefont {Hallmark}, \citenamefont
  {Droopad}, \citenamefont {Ooms}, \citenamefont {Salem}, \citenamefont
  {Bradshaw},\ and\ \citenamefont {Overgaard}}]{ApplPhysLett.76.1324}%
  \BibitemOpen
  \bibfield  {author} {\bibinfo {author} {\bibfnamefont {K.}~\bibnamefont
  {Eisenbeiser}}, \bibinfo {author} {\bibfnamefont {J.~M.}\ \bibnamefont
  {Finder}}, \bibinfo {author} {\bibfnamefont {Z.}~\bibnamefont {Yu}}, \bibinfo
  {author} {\bibfnamefont {J.}~\bibnamefont {Ramdani}}, \bibinfo {author}
  {\bibfnamefont {J.~A.}\ \bibnamefont {Curless}}, \bibinfo {author}
  {\bibfnamefont {J.~A.}\ \bibnamefont {Hallmark}}, \bibinfo {author}
  {\bibfnamefont {R.}~\bibnamefont {Droopad}}, \bibinfo {author} {\bibfnamefont
  {W.~J.}\ \bibnamefont {Ooms}}, \bibinfo {author} {\bibfnamefont
  {L.}~\bibnamefont {Salem}}, \bibinfo {author} {\bibfnamefont
  {S.}~\bibnamefont {Bradshaw}},\ and\ \bibinfo {author} {\bibfnamefont
  {C.~D.}\ \bibnamefont {Overgaard}},\ }\bibfield  {title} {\bibinfo {title}
  {Field effect transistors with SrTiO$_3$ gate dielectric on Si},\ }\href
  {https://doi.org/10.1063/1.126023} {\bibfield  {journal} {\bibinfo  {journal}
  {Appl. Phys. Lett.}\ }\textbf {\bibinfo {volume} {76}},\ \bibinfo {pages}
  {1324} (\bibinfo {year} {2000})}\BibitemShut {NoStop}%
\bibitem [{\citenamefont {Li}\ \emph {et~al.}(2003)\citenamefont {Li},
  \citenamefont {Li}, \citenamefont {Liu}, \citenamefont {Alim},\ and\
  \citenamefont {Chen}}]{JMaterSciMaterElectron.14.483}%
  \BibitemOpen
  \bibfield  {author} {\bibinfo {author} {\bibfnamefont {J.}~\bibnamefont
  {Li}}, \bibinfo {author} {\bibfnamefont {S.}~\bibnamefont {Li}}, \bibinfo
  {author} {\bibfnamefont {F.}~\bibnamefont {Liu}}, \bibinfo {author}
  {\bibfnamefont {M.~A.}\ \bibnamefont {Alim}},\ and\ \bibinfo {author}
  {\bibfnamefont {G.}~\bibnamefont {Chen}},\ }\bibfield  {title} {\bibinfo
  {title} {The origin of varistor property of SrTiO$_3$-based ceramics},\
  }\href {https://doi.org/10.1023/A:1023916716329} {\bibfield  {journal}
  {\bibinfo  {journal} {J. Mater. Sci.: Mater. Electron.}\ }\textbf {\bibinfo
  {volume} {14}},\ \bibinfo {pages} {483} (\bibinfo {year} {2003})}\BibitemShut
  {NoStop}%
\bibitem [{\citenamefont {Ohta}\ \emph {et~al.}(2005)\citenamefont {Ohta},
  \citenamefont {Nomura}, \citenamefont {Ohta}, \citenamefont {Hirano},
  \citenamefont {Hosono},\ and\ \citenamefont
  {Koumoto}}]{ApplPhysLett.87.092108}%
  \BibitemOpen
  \bibfield  {author} {\bibinfo {author} {\bibfnamefont {S.}~\bibnamefont
  {Ohta}}, \bibinfo {author} {\bibfnamefont {T.}~\bibnamefont {Nomura}},
  \bibinfo {author} {\bibfnamefont {H.}~\bibnamefont {Ohta}}, \bibinfo {author}
  {\bibfnamefont {M.}~\bibnamefont {Hirano}}, \bibinfo {author} {\bibfnamefont
  {H.}~\bibnamefont {Hosono}},\ and\ \bibinfo {author} {\bibfnamefont
  {K.}~\bibnamefont {Koumoto}},\ }\bibfield  {title} {\bibinfo {title} {Large
  thermoelectric performance of heavily Nb-doped SrTiO$_3$ epitaxial film at
  high temperature},\ }\href {https://doi.org/10.1063/1.2035889} {\bibfield
  {journal} {\bibinfo  {journal} {Appl. Phys. Lett.}\ }\textbf {\bibinfo
  {volume} {87}},\ \bibinfo {pages} {092108} (\bibinfo {year}
  {2005})}\BibitemShut {NoStop}%
\bibitem [{\citenamefont {Lin}\ \emph {et~al.}(2014)\citenamefont {Lin},
  \citenamefont {Bridoux}, \citenamefont {Gourgout}, \citenamefont {Seyfarth},
  \citenamefont {Kr\"amer}, \citenamefont {Nardone}, \citenamefont {Fauqu\'e},\
  and\ \citenamefont {Behnia}}]{PhysRevLett.112.207002}%
  \BibitemOpen
  \bibfield  {author} {\bibinfo {author} {\bibfnamefont {X.}~\bibnamefont
  {Lin}}, \bibinfo {author} {\bibfnamefont {G.}~\bibnamefont {Bridoux}},
  \bibinfo {author} {\bibfnamefont {A.}~\bibnamefont {Gourgout}}, \bibinfo
  {author} {\bibfnamefont {G.}~\bibnamefont {Seyfarth}}, \bibinfo {author}
  {\bibfnamefont {S.}~\bibnamefont {Kr\"amer}}, \bibinfo {author}
  {\bibfnamefont {M.}~\bibnamefont {Nardone}}, \bibinfo {author} {\bibfnamefont
  {B.}~\bibnamefont {Fauqu\'e}},\ and\ \bibinfo {author} {\bibfnamefont
  {K.}~\bibnamefont {Behnia}},\ }\bibfield  {title} {\bibinfo {title} {Critical
  doping for the onset of a two-band superconducting ground state in
  SrTiO$_{3-\delta}$},\ }\href {https://doi.org/10.1103/PhysRevLett.112.207002}
  {\bibfield  {journal} {\bibinfo  {journal} {Phys. Rev. Lett.}\ }\textbf
  {\bibinfo {volume} {112}},\ \bibinfo {pages} {207002} (\bibinfo {year}
  {2014})}\BibitemShut {NoStop}%
\bibitem [{\citenamefont {Park}\ \emph {et~al.}(2014)\citenamefont {Park},
  \citenamefont {Kwon}, \citenamefont {Park}, \citenamefont {Jung},\ and\
  \citenamefont {Kim}}]{ApplPhysLett.105.183103}%
  \BibitemOpen
  \bibfield  {author} {\bibinfo {author} {\bibfnamefont {J.}~\bibnamefont
  {Park}}, \bibinfo {author} {\bibfnamefont {D.-H.}\ \bibnamefont {Kwon}},
  \bibinfo {author} {\bibfnamefont {H.}~\bibnamefont {Park}}, \bibinfo {author}
  {\bibfnamefont {C.~U.}\ \bibnamefont {Jung}},\ and\ \bibinfo {author}
  {\bibfnamefont {M.}~\bibnamefont {Kim}},\ }\bibfield  {title} {\bibinfo
  {title} {Role of oxygen vacancies in resistive switching in Pt/Nb-doped
  SrTiO$_3$},\ }\href {https://doi.org/10.1063/1.4901053} {\bibfield  {journal}
  {\bibinfo  {journal} {Appl. Phys. Lett.}\ }\textbf {\bibinfo {volume}
  {105}},\ \bibinfo {pages} {183103} (\bibinfo {year} {2014})}\BibitemShut
  {NoStop}%
\bibitem [{\citenamefont {Kato}\ and\ \citenamefont
  {Kudo}(2002)}]{JPhysChemB.106.5029}%
  \BibitemOpen
  \bibfield  {author} {\bibinfo {author} {\bibfnamefont {H.}~\bibnamefont
  {Kato}}\ and\ \bibinfo {author} {\bibfnamefont {A.}~\bibnamefont {Kudo}},\
  }\bibfield  {title} {\bibinfo {title} {Visible-light-response and
  photocatalytic activities of TiO$_2$ and SrTiO$_3$ photocatalysts codoped
  with antimony and chromium},\ }\href {https://doi.org/10.1021/jp0255482}
  {\bibfield  {journal} {\bibinfo  {journal} {J. Phys. Chem. B}\ }\textbf
  {\bibinfo {volume} {106}},\ \bibinfo {pages} {5029} (\bibinfo {year}
  {2002})}\BibitemShut {NoStop}%
\bibitem [{\citenamefont {Konta}\ \emph {et~al.}(2004)\citenamefont {Konta},
  \citenamefont {Ishii}, \citenamefont {Kato},\ and\ \citenamefont
  {Kudo}}]{JPhysChemB.108.8992}%
  \BibitemOpen
  \bibfield  {author} {\bibinfo {author} {\bibfnamefont {R.}~\bibnamefont
  {Konta}}, \bibinfo {author} {\bibfnamefont {T.}~\bibnamefont {Ishii}},
  \bibinfo {author} {\bibfnamefont {H.}~\bibnamefont {Kato}},\ and\ \bibinfo
  {author} {\bibfnamefont {A.}~\bibnamefont {Kudo}},\ }\bibfield  {title}
  {\bibinfo {title} {Photocatalytic activities of noble metal ion doped
  SrTiO$_3$ under visible light irradiation},\ }\href
  {https://doi.org/10.1021/jp049556p} {\bibfield  {journal} {\bibinfo
  {journal} {J. Phys. Chem. B}\ }\textbf {\bibinfo {volume} {108}},\ \bibinfo
  {pages} {8992} (\bibinfo {year} {2004})}\BibitemShut {NoStop}%
\bibitem [{\citenamefont {Hara}\ and\ \citenamefont
  {Ishiguro}(2009)}]{SensActuatorsB.136.489}%
  \BibitemOpen
  \bibfield  {author} {\bibinfo {author} {\bibfnamefont {T.}~\bibnamefont
  {Hara}}\ and\ \bibinfo {author} {\bibfnamefont {T.}~\bibnamefont
  {Ishiguro}},\ }\bibfield  {title} {\bibinfo {title} {Oxygen sensitivity of
  SrTiO$_3$ thin film prepared using atomic layer deposition},\ }\href
  {https://doi.org/10.1016/j.snb.2008.12.026} {\bibfield  {journal} {\bibinfo
  {journal} {Sens. Actuators B}\ }\textbf {\bibinfo {volume} {136}},\ \bibinfo
  {pages} {489} (\bibinfo {year} {2009})}\BibitemShut {NoStop}%
\bibitem [{\citenamefont {Lemanov}\ \emph {et~al.}(2004)\citenamefont
  {Lemanov}, \citenamefont {Smirnova}, \citenamefont {Sotnikov},\ and\
  \citenamefont {Weihnacht}}]{PhysSolidState.46.1442}%
  \BibitemOpen
  \bibfield  {author} {\bibinfo {author} {\bibfnamefont {V.~V.}\ \bibnamefont
  {Lemanov}}, \bibinfo {author} {\bibfnamefont {E.~P.}\ \bibnamefont
  {Smirnova}}, \bibinfo {author} {\bibfnamefont {A.~V.}\ \bibnamefont
  {Sotnikov}},\ and\ \bibinfo {author} {\bibfnamefont {M.}~\bibnamefont
  {Weihnacht}},\ }\bibfield  {title} {\bibinfo {title} {Dielectric relaxation
  in SrTiO$_3$:Mn},\ }\href {https://doi.org/10.1134/1.1788776} {\bibfield
  {journal} {\bibinfo  {journal} {Phys. Solid State}\ }\textbf {\bibinfo
  {volume} {46}},\ \bibinfo {pages} {1442} (\bibinfo {year}
  {2004})}\BibitemShut {NoStop}%
\bibitem [{\citenamefont {Lemanov}\ \emph {et~al.}(2005)\citenamefont
  {Lemanov}, \citenamefont {Sotnikov}, \citenamefont {Smirnova},\ and\
  \citenamefont {Weihnacht}}]{JApplPhys.98.056102}%
  \BibitemOpen
  \bibfield  {author} {\bibinfo {author} {\bibfnamefont {V.~V.}\ \bibnamefont
  {Lemanov}}, \bibinfo {author} {\bibfnamefont {A.~V.}\ \bibnamefont
  {Sotnikov}}, \bibinfo {author} {\bibfnamefont {E.~P.}\ \bibnamefont
  {Smirnova}},\ and\ \bibinfo {author} {\bibfnamefont {M.}~\bibnamefont
  {Weihnacht}},\ }\bibfield  {title} {\bibinfo {title} {Dielectric relaxation
  in doped SrTiO$_3$: Transition from classical thermal activation to quantum
  tunnelling},\ }\href {https://doi.org/10.1063/1.2035313} {\bibfield
  {journal} {\bibinfo  {journal} {J. Appl. Phys.}\ }\textbf {\bibinfo {volume}
  {98}},\ \bibinfo {pages} {056102} (\bibinfo {year} {2005})}\BibitemShut
  {NoStop}%
\bibitem [{\citenamefont {Tkach}\ \emph {et~al.}(2005)\citenamefont {Tkach},
  \citenamefont {Vilarinho},\ and\ \citenamefont
  {Kholkin}}]{ApplPhysLett.86.172902}%
  \BibitemOpen
  \bibfield  {author} {\bibinfo {author} {\bibfnamefont {A.}~\bibnamefont
  {Tkach}}, \bibinfo {author} {\bibfnamefont {P.~M.}\ \bibnamefont
  {Vilarinho}},\ and\ \bibinfo {author} {\bibfnamefont {A.~L.}\ \bibnamefont
  {Kholkin}},\ }\bibfield  {title} {\bibinfo {title} {Polar behavior in
  Mn-doped SrTiO$_3$ ceramics},\ }\href {https://doi.org/10.1063/1.1920414}
  {\bibfield  {journal} {\bibinfo  {journal} {Appl. Phys. Lett.}\ }\textbf
  {\bibinfo {volume} {86}},\ \bibinfo {pages} {172902} (\bibinfo {year}
  {2005})}\BibitemShut {NoStop}%
\bibitem [{\citenamefont {Lebedev}\ \emph {et~al.}(2009)\citenamefont
  {Lebedev}, \citenamefont {Sluchinskaya}, \citenamefont {Erko},\ and\
  \citenamefont {Kozlovskii}}]{JETPLett.89.457}%
  \BibitemOpen
  \bibfield  {author} {\bibinfo {author} {\bibfnamefont {A.~I.}\ \bibnamefont
  {Lebedev}}, \bibinfo {author} {\bibfnamefont {I.~A.}\ \bibnamefont
  {Sluchinskaya}}, \bibinfo {author} {\bibfnamefont {A.}~\bibnamefont {Erko}},\
  and\ \bibinfo {author} {\bibfnamefont {V.~F.}\ \bibnamefont {Kozlovskii}},\
  }\bibfield  {title} {\bibinfo {title} {Direct evidence for off-centering of
  Mn impurity in SrTiO$_3$},\ }\href
  {https://doi.org/10.1134/S0021364009090070} {\bibfield  {journal} {\bibinfo
  {journal} {JETP Lett.}\ }\textbf {\bibinfo {volume} {89}},\ \bibinfo {pages}
  {457} (\bibinfo {year} {2009})}\BibitemShut {NoStop}%
\bibitem [{\citenamefont {Levin}\ \emph {et~al.}(2010)\citenamefont {Levin},
  \citenamefont {Krayzman}, \citenamefont {Woicik}, \citenamefont {Tkach},\
  and\ \citenamefont {Vilarinho}}]{ApplPhysLett.96.052904}%
  \BibitemOpen
  \bibfield  {author} {\bibinfo {author} {\bibfnamefont {I.}~\bibnamefont
  {Levin}}, \bibinfo {author} {\bibfnamefont {V.}~\bibnamefont {Krayzman}},
  \bibinfo {author} {\bibfnamefont {J.~C.}\ \bibnamefont {Woicik}}, \bibinfo
  {author} {\bibfnamefont {A.}~\bibnamefont {Tkach}},\ and\ \bibinfo {author}
  {\bibfnamefont {P.~M.}\ \bibnamefont {Vilarinho}},\ }\bibfield  {title}
  {\bibinfo {title} {X-ray absorption fine structure studies of Mn coordination
  in doped perovskite SrTiO$_3$},\ }\href {https://doi.org/10.1063/1.3298369}
  {\bibfield  {journal} {\bibinfo  {journal} {Appl. Phys. Lett.}\ }\textbf
  {\bibinfo {volume} {96}},\ \bibinfo {pages} {052904} (\bibinfo {year}
  {2010})}\BibitemShut {NoStop}%
\bibitem [{\citenamefont {Sluchinskaya}\ \emph {et~al.}(2010)\citenamefont
  {Sluchinskaya}, \citenamefont {Lebedev},\ and\ \citenamefont
  {Erko}}]{BullRASPhys.74.1235}%
  \BibitemOpen
  \bibfield  {author} {\bibinfo {author} {\bibfnamefont {I.~A.}\ \bibnamefont
  {Sluchinskaya}}, \bibinfo {author} {\bibfnamefont {A.~I.}\ \bibnamefont
  {Lebedev}},\ and\ \bibinfo {author} {\bibfnamefont {A.}~\bibnamefont
  {Erko}},\ }\bibfield  {title} {\bibinfo {title} {Local environment and
  oxidation state of a Mn impurity in SrTiO$_3$ determined from XAFS data},\
  }\href {https://doi.org/10.3103/S1062873810090145} {\bibfield  {journal}
  {\bibinfo  {journal} {Bull. Russ. Acad. Sci.: Phys.}\ }\textbf {\bibinfo
  {volume} {74}},\ \bibinfo {pages} {1235} (\bibinfo {year}
  {2010})}\BibitemShut {NoStop}%
\bibitem [{\citenamefont {Valant}\ \emph {et~al.}(2012)\citenamefont {Valant},
  \citenamefont {Kolodiazhnyi}, \citenamefont {Ar{\v{c}}on}, \citenamefont
  {Aguesse}, \citenamefont {Axelsson},\ and\ \citenamefont
  {Alford}}]{AdvFunctMater.22.2114}%
  \BibitemOpen
  \bibfield  {author} {\bibinfo {author} {\bibfnamefont {M.}~\bibnamefont
  {Valant}}, \bibinfo {author} {\bibfnamefont {T.}~\bibnamefont
  {Kolodiazhnyi}}, \bibinfo {author} {\bibfnamefont {I.}~\bibnamefont
  {Ar{\v{c}}on}}, \bibinfo {author} {\bibfnamefont {F.}~\bibnamefont
  {Aguesse}}, \bibinfo {author} {\bibfnamefont {A.-K.}\ \bibnamefont
  {Axelsson}},\ and\ \bibinfo {author} {\bibfnamefont {N.~M.}\ \bibnamefont
  {Alford}},\ }\bibfield  {title} {\bibinfo {title} {The origin of magnetism in
  Mn-doped SrTiO$_3$},\ }\href {https://doi.org/10.1002/adfm.201102482}
  {\bibfield  {journal} {\bibinfo  {journal} {Adv. Funct. Mater.}\ }\textbf
  {\bibinfo {volume} {22}},\ \bibinfo {pages} {2114} (\bibinfo {year}
  {2012})}\BibitemShut {NoStop}%
\bibitem [{\citenamefont {Maier}\ \emph {et~al.}(2020)\citenamefont {Maier},
  \citenamefont {Cockayne}, \citenamefont {Donohue}, \citenamefont {Cibin},\
  and\ \citenamefont {Levin}}]{ChemMater.32.4651}%
  \BibitemOpen
  \bibfield  {author} {\bibinfo {author} {\bibfnamefont {R.~A.}\ \bibnamefont
  {Maier}}, \bibinfo {author} {\bibfnamefont {E.}~\bibnamefont {Cockayne}},
  \bibinfo {author} {\bibfnamefont {M.}~\bibnamefont {Donohue}}, \bibinfo
  {author} {\bibfnamefont {G.}~\bibnamefont {Cibin}},\ and\ \bibinfo {author}
  {\bibfnamefont {I.}~\bibnamefont {Levin}},\ }\bibfield  {title} {\bibinfo
  {title} {Substitutional mechanisms and structural relaxations for manganese
  in SrTiO$_3$: Bridging the concentration gap for point-defect metrology},\
  }\href {https://doi.org/10.1021/acs.chemmater.0c01082} {\bibfield  {journal}
  {\bibinfo  {journal} {Chem. Mater.}\ }\textbf {\bibinfo {volume} {32}},\
  \bibinfo {pages} {4651} (\bibinfo {year} {2020})}\BibitemShut {NoStop}%
\bibitem [{\citenamefont {Shvartsman}\ \emph {et~al.}(2008)\citenamefont
  {Shvartsman}, \citenamefont {Bedanta}, \citenamefont {Borisov}, \citenamefont
  {Kleemann}, \citenamefont {Tkach},\ and\ \citenamefont
  {Vilarinho}}]{PhysRevLett.101.165704}%
  \BibitemOpen
  \bibfield  {author} {\bibinfo {author} {\bibfnamefont {V.~V.}\ \bibnamefont
  {Shvartsman}}, \bibinfo {author} {\bibfnamefont {S.}~\bibnamefont {Bedanta}},
  \bibinfo {author} {\bibfnamefont {P.}~\bibnamefont {Borisov}}, \bibinfo
  {author} {\bibfnamefont {W.}~\bibnamefont {Kleemann}}, \bibinfo {author}
  {\bibfnamefont {A.}~\bibnamefont {Tkach}},\ and\ \bibinfo {author}
  {\bibfnamefont {P.~M.}\ \bibnamefont {Vilarinho}},\ }\bibfield  {title}
  {\bibinfo {title} {(Sr,Mn)TiO$_3$: A magnetoelectric multiglass},\ }\href
  {https://doi.org/10.1103/PhysRevLett.101.165704} {\bibfield  {journal}
  {\bibinfo  {journal} {Phys. Rev. Lett.}\ }\textbf {\bibinfo {volume} {101}},\
  \bibinfo {pages} {165704} (\bibinfo {year} {2008})}\BibitemShut {NoStop}%
\bibitem [{\citenamefont {Kleemann}\ \emph {et~al.}(2008)\citenamefont
  {Kleemann}, \citenamefont {Shvartsman}, \citenamefont {Bedanta},
  \citenamefont {Borisov}, \citenamefont {Tkach},\ and\ \citenamefont
  {Vilarinho}}]{JPhysCondensMatter.20.434216}%
  \BibitemOpen
  \bibfield  {author} {\bibinfo {author} {\bibfnamefont {W.}~\bibnamefont
  {Kleemann}}, \bibinfo {author} {\bibfnamefont {V.~V.}\ \bibnamefont
  {Shvartsman}}, \bibinfo {author} {\bibfnamefont {S.}~\bibnamefont {Bedanta}},
  \bibinfo {author} {\bibfnamefont {P.}~\bibnamefont {Borisov}}, \bibinfo
  {author} {\bibfnamefont {A.}~\bibnamefont {Tkach}},\ and\ \bibinfo {author}
  {\bibfnamefont {P.~M.}\ \bibnamefont {Vilarinho}},\ }\bibfield  {title}
  {\bibinfo {title} {(Sr,Mn)TiO$_3$---a magnetoelectrically coupled
  multiglass},\ }\href {https://doi.org/10.1088/0953-8984/20/43/434216}
  {\bibfield  {journal} {\bibinfo  {journal} {J. Phys.: Condens. Matter}\
  }\textbf {\bibinfo {volume} {20}},\ \bibinfo {pages} {434216} (\bibinfo
  {year} {2008})}\BibitemShut {NoStop}%
\bibitem [{\citenamefont {Sluchinskaya}\ and\ \citenamefont
  {Lebedev}(2019)}]{PhysSolidState.61.390}%
  \BibitemOpen
  \bibfield  {author} {\bibinfo {author} {\bibfnamefont {I.~A.}\ \bibnamefont
  {Sluchinskaya}}\ and\ \bibinfo {author} {\bibfnamefont {A.~I.}\ \bibnamefont
  {Lebedev}},\ }\bibfield  {title} {\bibinfo {title} {Cobalt in strontium
  titanate as a new off-center magnetic impurity},\ }\href
  {https://doi.org/10.1134/S1063783419030302} {\bibfield  {journal} {\bibinfo
  {journal} {Phys. Solid State}\ }\textbf {\bibinfo {volume} {61}},\ \bibinfo
  {pages} {390} (\bibinfo {year} {2019})}\BibitemShut {NoStop}%
\bibitem [{\citenamefont {Sluchinskaya}\ and\ \citenamefont
  {Lebedev}(2020)}]{JAlloysComp.820.153243}%
  \BibitemOpen
  \bibfield  {author} {\bibinfo {author} {\bibfnamefont {I.~A.}\ \bibnamefont
  {Sluchinskaya}}\ and\ \bibinfo {author} {\bibfnamefont {A.~I.}\ \bibnamefont
  {Lebedev}},\ }\bibfield  {title} {\bibinfo {title} {Electronic and magnetic
  properties of structural defects in SrTiO$_3$(Co)},\ }\href
  {https://doi.org/10.1016/j.jallcom.2019.153243} {\bibfield  {journal}
  {\bibinfo  {journal} {J. Alloys Comp.}\ }\textbf {\bibinfo {volume} {820}},\
  \bibinfo {pages} {153243} (\bibinfo {year} {2020})}\BibitemShut {NoStop}%
\bibitem [{\citenamefont {Garg}\ \emph {et~al.}(2018)\citenamefont {Garg},
  \citenamefont {Kumar},\ and\ \citenamefont {Nair}}]{PhysRevMater.2.094409}%
  \BibitemOpen
  \bibfield  {author} {\bibinfo {author} {\bibfnamefont {C.}~\bibnamefont
  {Garg}}, \bibinfo {author} {\bibfnamefont {J.}~\bibnamefont {Kumar}},\ and\
  \bibinfo {author} {\bibfnamefont {S.}~\bibnamefont {Nair}},\ }\bibfield
  {title} {\bibinfo {title} {Absence of a multiglass state in some transition
  metal doped quantum paraelectrics},\ }\href
  {https://doi.org/10.1103/PhysRevMaterials.2.094409} {\bibfield  {journal}
  {\bibinfo  {journal} {Phys. Rev. Mater.}\ }\textbf {\bibinfo {volume} {2}},\
  \bibinfo {pages} {094409} (\bibinfo {year} {2018})}\BibitemShut {NoStop}%
\bibitem [{\citenamefont {Molin}\ \emph {et~al.}(2012)\citenamefont {Molin},
  \citenamefont {Lewandowska-Iwaniak}, \citenamefont {Kusz}, \citenamefont
  {Gazda},\ and\ \citenamefont {Jasinski}}]{JElectroceram.28.80}%
  \BibitemOpen
  \bibfield  {author} {\bibinfo {author} {\bibfnamefont {S.}~\bibnamefont
  {Molin}}, \bibinfo {author} {\bibfnamefont {W.}~\bibnamefont
  {Lewandowska-Iwaniak}}, \bibinfo {author} {\bibfnamefont {B.}~\bibnamefont
  {Kusz}}, \bibinfo {author} {\bibfnamefont {M.}~\bibnamefont {Gazda}},\ and\
  \bibinfo {author} {\bibfnamefont {P.}~\bibnamefont {Jasinski}},\ }\bibfield
  {title} {\bibinfo {title} {Structural and electrical properties of Sr(Ti,
  Fe)O$_{3-\delta}$ materials for SOFC cathodes},\ }\href
  {https://doi.org/10.1007/s10832-012-9683-x} {\bibfield  {journal} {\bibinfo
  {journal} {J. Electroceramics}\ }\textbf {\bibinfo {volume} {28}},\ \bibinfo
  {pages} {80} (\bibinfo {year} {2012})}\BibitemShut {NoStop}%
\bibitem [{\citenamefont {Faughnan}(1971)}]{PhysRevB.4.3623}%
  \BibitemOpen
  \bibfield  {author} {\bibinfo {author} {\bibfnamefont {B.~W.}\ \bibnamefont
  {Faughnan}},\ }\bibfield  {title} {\bibinfo {title} {Photochromism in
  transition-metal-doped SrTiO$_3$},\ }\href
  {https://doi.org/10.1103/PhysRevB.4.3623} {\bibfield  {journal} {\bibinfo
  {journal} {Phys. Rev. B}\ }\textbf {\bibinfo {volume} {4}},\ \bibinfo {pages}
  {3623} (\bibinfo {year} {1971})}\BibitemShut {NoStop}%
\bibitem [{\citenamefont {Lenser}\ \emph {et~al.}(2011)\citenamefont {Lenser},
  \citenamefont {Kalinko}, \citenamefont {Kuzmin}, \citenamefont {Berzins},
  \citenamefont {Purans}, \citenamefont {Szot}, \citenamefont {Waser},\ and\
  \citenamefont {Dittmann}}]{PhysChemChemPhys.13.20779}%
  \BibitemOpen
  \bibfield  {author} {\bibinfo {author} {\bibfnamefont {C.}~\bibnamefont
  {Lenser}}, \bibinfo {author} {\bibfnamefont {A.}~\bibnamefont {Kalinko}},
  \bibinfo {author} {\bibfnamefont {A.}~\bibnamefont {Kuzmin}}, \bibinfo
  {author} {\bibfnamefont {D.}~\bibnamefont {Berzins}}, \bibinfo {author}
  {\bibfnamefont {J.}~\bibnamefont {Purans}}, \bibinfo {author} {\bibfnamefont
  {K.}~\bibnamefont {Szot}}, \bibinfo {author} {\bibfnamefont {R.}~\bibnamefont
  {Waser}},\ and\ \bibinfo {author} {\bibfnamefont {R.}~\bibnamefont
  {Dittmann}},\ }\bibfield  {title} {\bibinfo {title} {Spectroscopic study of
  the electric field induced valence change of Fe-defect centers in
  SrTiO$_3$},\ }\href {https://doi.org/10.1039/C1CP21973A} {\bibfield
  {journal} {\bibinfo  {journal} {Phys. Chem. Chem. Phys.}\ }\textbf {\bibinfo
  {volume} {13}},\ \bibinfo {pages} {20779} (\bibinfo {year}
  {2011})}\BibitemShut {NoStop}%
\bibitem [{\citenamefont {Vra{\v{c}}ar}\ \emph {et~al.}(2007)\citenamefont
  {Vra{\v{c}}ar}, \citenamefont {Kuzmin}, \citenamefont {Merkle}, \citenamefont
  {Purans}, \citenamefont {Kotomin}, \citenamefont {Maier},\ and\ \citenamefont
  {Mathon}}]{PhysRevB.76.174107}%
  \BibitemOpen
  \bibfield  {author} {\bibinfo {author} {\bibfnamefont {M.}~\bibnamefont
  {Vra{\v{c}}ar}}, \bibinfo {author} {\bibfnamefont {A.}~\bibnamefont
  {Kuzmin}}, \bibinfo {author} {\bibfnamefont {R.}~\bibnamefont {Merkle}},
  \bibinfo {author} {\bibfnamefont {J.}~\bibnamefont {Purans}}, \bibinfo
  {author} {\bibfnamefont {E.~A.}\ \bibnamefont {Kotomin}}, \bibinfo {author}
  {\bibfnamefont {J.}~\bibnamefont {Maier}},\ and\ \bibinfo {author}
  {\bibfnamefont {O.}~\bibnamefont {Mathon}},\ }\bibfield  {title} {\bibinfo
  {title} {Jahn-Teller distortion around Fe$^{4+}$ in
  Sr(Fe$_x$Ti$_{1-x}$)O$_{3-\delta}$ from X-ray absorption spectroscopy, X-ray
  diffraction, and vibrational spectroscopy},\ }\href
  {https://doi.org/10.1103/PhysRevB.76.174107} {\bibfield  {journal} {\bibinfo
  {journal} {Phys. Rev. B}\ }\textbf {\bibinfo {volume} {76}},\ \bibinfo
  {pages} {174107} (\bibinfo {year} {2007})}\BibitemShut {NoStop}%
\bibitem [{\citenamefont {Alexandrov}\ \emph {et~al.}(2008)\citenamefont
  {Alexandrov}, \citenamefont {Maier},\ and\ \citenamefont
  {Evarestov}}]{PhysRevB.77.075111}%
  \BibitemOpen
  \bibfield  {author} {\bibinfo {author} {\bibfnamefont {V.~E.}\ \bibnamefont
  {Alexandrov}}, \bibinfo {author} {\bibfnamefont {J.}~\bibnamefont {Maier}},\
  and\ \bibinfo {author} {\bibfnamefont {R.~A.}\ \bibnamefont {Evarestov}},\
  }\bibfield  {title} {\bibinfo {title} {\emph{Ab initio} study of
  SrFe$_x$Ti$_{1-x}$O$_3$: Jahn-Teller distortion and electronic structure},\
  }\href {https://doi.org/10.1103/PhysRevB.77.075111} {\bibfield  {journal}
  {\bibinfo  {journal} {Phys. Rev. B}\ }\textbf {\bibinfo {volume} {77}},\
  \bibinfo {pages} {075111} (\bibinfo {year} {2008})}\BibitemShut {NoStop}%
\bibitem [{\citenamefont {Koehl}\ \emph {et~al.}(2013)\citenamefont {Koehl},
  \citenamefont {Kajewski}, \citenamefont {Kubacki}, \citenamefont {Lenser},
  \citenamefont {Dittmann}, \citenamefont {Meuffels}, \citenamefont {Szot},
  \citenamefont {Waser},\ and\ \citenamefont
  {Szade}}]{PhysChemChemPhys.15.8311}%
  \BibitemOpen
  \bibfield  {author} {\bibinfo {author} {\bibfnamefont {A.}~\bibnamefont
  {Koehl}}, \bibinfo {author} {\bibfnamefont {D.}~\bibnamefont {Kajewski}},
  \bibinfo {author} {\bibfnamefont {J.}~\bibnamefont {Kubacki}}, \bibinfo
  {author} {\bibfnamefont {C.}~\bibnamefont {Lenser}}, \bibinfo {author}
  {\bibfnamefont {R.}~\bibnamefont {Dittmann}}, \bibinfo {author}
  {\bibfnamefont {P.}~\bibnamefont {Meuffels}}, \bibinfo {author}
  {\bibfnamefont {K.}~\bibnamefont {Szot}}, \bibinfo {author} {\bibfnamefont
  {R.}~\bibnamefont {Waser}},\ and\ \bibinfo {author} {\bibfnamefont
  {J.}~\bibnamefont {Szade}},\ }\bibfield  {title} {\bibinfo {title} {Detection
  of Fe$^{2+}$ valence states in Fe doped SrTiO$_3$ epitaxial thin films grown
  by pulsed laser deposition},\ }\href {https://doi.org/10.1039/C3CP50272D}
  {\bibfield  {journal} {\bibinfo  {journal} {Phys. Chem. Chem. Phys.}\
  }\textbf {\bibinfo {volume} {15}},\ \bibinfo {pages} {8311} (\bibinfo {year}
  {2013})}\BibitemShut {NoStop}%
\bibitem [{\citenamefont {Lebedev}\ \emph {et~al.}(1996)\citenamefont
  {Lebedev}, \citenamefont {Sluchinskaya}, \citenamefont {Demin},\ and\
  \citenamefont {H.}}]{BullRASPhys.60.1533}%
  \BibitemOpen
  \bibfield  {author} {\bibinfo {author} {\bibfnamefont {A.~I.}\ \bibnamefont
  {Lebedev}}, \bibinfo {author} {\bibfnamefont {I.~A.}\ \bibnamefont
  {Sluchinskaya}}, \bibinfo {author} {\bibfnamefont {V.~N.}\ \bibnamefont
  {Demin}},\ and\ \bibinfo {author} {\bibfnamefont {I.~H.}\ \bibnamefont
  {Munro}},\ }\bibfield  {title} {\bibinfo {title} {Impurity effect on the phase
  transition in GeTe studied by the EXAFS spectroscopy},\ }\href@noop {}
  {\bibfield  {journal} {\bibinfo  {journal} {Bull. Russ. Akad. Sci.: Phys.}\
  }\textbf {\bibinfo {volume} {60}},\ \bibinfo {pages} {1533} (\bibinfo {year}
  {1996})}\BibitemShut {NoStop}%
\bibitem [{\citenamefont {Lebedev}\ \emph {et~al.}(1997)\citenamefont
  {Lebedev}, \citenamefont {Sluchinskaya}, \citenamefont {Demin},\ and\
  \citenamefont {Munro}}]{PhysRevB.55.14770}%
  \BibitemOpen
  \bibfield  {author} {\bibinfo {author} {\bibfnamefont {A.~I.}\ \bibnamefont
  {Lebedev}}, \bibinfo {author} {\bibfnamefont {I.~A.}\ \bibnamefont
  {Sluchinskaya}}, \bibinfo {author} {\bibfnamefont {V.~N.}\ \bibnamefont
  {Demin}},\ and\ \bibinfo {author} {\bibfnamefont {I.~H.}\ \bibnamefont
  {Munro}},\ }\bibfield  {title} {\bibinfo {title} {Off-centering of Pb and Sn
  impurities in GeTe},\ }\href {https://doi.org/10.1103/PhysRevB.55.14770}
  {\bibfield  {journal} {\bibinfo  {journal} {Phys. Rev. B}\ }\textbf {\bibinfo
  {volume} {55}},\ \bibinfo {pages} {14770} (\bibinfo {year}
  {1997})}\BibitemShut {NoStop}%
\bibitem [{\citenamefont {Ravel}\ and\ \citenamefont
  {Newville}(2005)}]{JSynchrotronRad.12.537}%
  \BibitemOpen
  \bibfield  {author} {\bibinfo {author} {\bibfnamefont {B.}~\bibnamefont
  {Ravel}}\ and\ \bibinfo {author} {\bibfnamefont {M.}~\bibnamefont
  {Newville}},\ }\bibfield  {title} {\bibinfo {title} {{\it ATHENA}, {\it
  ARTEMIS}, {\it HEPHAESTUS}: data analysis for X-ray absorption spectroscopy
  using {\it IFEFFIT}},\ }\href {https://doi.org/10.1107/S0909049505012719}
  {\bibfield  {journal} {\bibinfo  {journal} {J. Synchrotron Rad.}\ }\textbf
  {\bibinfo {volume} {12}},\ \bibinfo {pages} {537} (\bibinfo {year}
  {2005})}\BibitemShut {NoStop}%
\bibitem [{\citenamefont {Sluchinskaya}\ and\ \citenamefont
  {Lebedev}(2017)}]{PhysSolidState.59.1512}%
  \BibitemOpen
  \bibfield  {author} {\bibinfo {author} {\bibfnamefont {I.~A.}\ \bibnamefont
  {Sluchinskaya}}\ and\ \bibinfo {author} {\bibfnamefont {A.~I.}\ \bibnamefont
  {Lebedev}},\ }\bibfield  {title} {\bibinfo {title} {An experimental and
  theoretical study of Ni impurity centers in Ba$_{0.8}$Sr$_{0.2}$TiO$_3$},\
  }\href {https://doi.org/10.1134/S106378341708025X} {\bibfield  {journal}
  {\bibinfo  {journal} {Phys. Solid State}\ }\textbf {\bibinfo {volume} {59}},\
  \bibinfo {pages} {1512} (\bibinfo {year} {2017})}\BibitemShut {NoStop}%
\bibitem [{\citenamefont {Garrity}\ \emph {et~al.}(2014)\citenamefont
  {Garrity}, \citenamefont {Bennett}, \citenamefont {Rabe},\ and\ \citenamefont
  {Vanderbilt}}]{ComputMaterSci.81.446}%
  \BibitemOpen
  \bibfield  {author} {\bibinfo {author} {\bibfnamefont {K.~F.}\ \bibnamefont
  {Garrity}}, \bibinfo {author} {\bibfnamefont {J.~W.}\ \bibnamefont
  {Bennett}}, \bibinfo {author} {\bibfnamefont {K.~M.}\ \bibnamefont {Rabe}},\
  and\ \bibinfo {author} {\bibfnamefont {D.}~\bibnamefont {Vanderbilt}},\
  }\bibfield  {title} {\bibinfo {title} {Pseudopotentials for high throughput
  DFT calculations},\ }\href {https://doi.org/10.1016/j.commatsci.2013.08.053}
  {\bibfield  {journal} {\bibinfo  {journal} {Comput. Mater. Sci.}\ }\textbf
  {\bibinfo {volume} {81}},\ \bibinfo {pages} {446} (\bibinfo {year}
  {2014})}\BibitemShut {NoStop}%
\bibitem [{\citenamefont {Lebedev}\ and\ \citenamefont
  {Sluchinskaya}(2016)}]{Ferroelectrics.501.1}%
  \BibitemOpen
  \bibfield  {author} {\bibinfo {author} {\bibfnamefont {A.~I.}\ \bibnamefont
  {Lebedev}}\ and\ \bibinfo {author} {\bibfnamefont {I.~A.}\ \bibnamefont
  {Sluchinskaya}},\ }\bibfield  {title} {\bibinfo {title} {On the nature of
  change in Ni oxidation state in BaTiO$_3$--SrTiO$_3$ system},\ }\href
  {https://doi.org/10.1080/00150193.2016.1198196} {\bibfield  {journal}
  {\bibinfo  {journal} {Ferroelectrics}\ }\textbf {\bibinfo {volume} {501}},\
  \bibinfo {pages} {1} (\bibinfo {year} {2016})}\BibitemShut {NoStop}%
\end{thebibliography}
\providecommand{\BIBYu}{Yu}

\end{document}